\documentclass[12pt]{article}
\usepackage{amsfonts}
\usepackage{cite}
\usepackage{amsmath}
\usepackage{amssymb}
\usepackage{graphicx}

\UseRawInputEncoding
\providecommand{\U}[1]{\protect\rule{.1in}{.1in}}
\makeatletter
\@addtoreset{equation}{section}
\makeatother

\input{tcilatex}
\begin{document}

%TCIMACRO{%
%\TeXButton{titlepage}{\begin{titlepage}
%\begin{flushright}
%\end{flushright}
%\vspace{.3cm}
%\begin{center}
%\renewcommand{\thefootnote}{\fnsymbol{footnote}}
%{\Large{\bf Kasner Metrics and Very Special Geometry}}
%\vskip1cm
%\vskip1.3cm
%W. A. Sabra 
%\vskip1cm
%{\small{\it
%Physics Department, 
%American University of Beirut\\ Lebanon  \\}}
%\end{center}
%\vfill\begin{center}
%\textbf{Abstract}
%\end{center}
%\begin{quote}
%We consider general charged Kasner-like  solutions for the theory of five-dimensional supergravity coupled to 
%Abelian vector multiplets in arbitrary space-time signature. These solutions, depending on the choice of coordinates, can be 
%thought of as generalisations of Melvin/Rosen cosmologies, flux-branes and domain walls. 
%\end{quote}
%\vfill\end{titlepage}}}%
%BeginExpansion
\begin{titlepage}
\begin{flushright}
\end{flushright}
\vspace{.3cm}
\begin{center}
\renewcommand{\thefootnote}{\fnsymbol{footnote}}
{\Large{\bf Kasner Metrics and Very Special Geometry}}
\vskip1cm
\vskip1.3cm
W. A. Sabra 
\vskip1cm
{\small{\it
Physics Department, 
American University of Beirut\\ Lebanon  \\}}
\end{center}
\vfill\begin{center}
\textbf{Abstract}
\end{center}
\begin{quote}
We consider general charged Kasner-like  solutions for the theory of five-dimensional supergravity coupled to 
Abelian vector multiplets in arbitrary space-time signature. These solutions, depending on the choice of coordinates, can be 
thought of as generalisations of Melvin/Rosen cosmologies, flux-branes and domain walls. 
\end{quote}
\vfill\end{titlepage}%
%EndExpansion

\section{Introduction}

In recent years an active area of research has been the classification and
analysis of supersymmetric gravitational backgrounds in supergravity
theories with various space-time dimensions and signatures. The present work
deals with non-supersymmetric cosmological and static solutions of the
theories of five-dimensional supergravities coupled to vector multiplets in
arbitrary space-time signature \cite{GST}. In these supergravity theories,
the dynamics of the bosonic fields can be described in terms of the so
called very special geometry \cite{speone}. Early analysis of supersymmetric
solutions, with vanishing fermionic fields, was based on starting with a
specific ansatz for the space-time metric and fixing it alongside the
bosonic fields by requiring the existence of some Killing spinors (see for
example \cite{sc}). A systematic classification of solutions admitting
Killing spinors for minimal and general five-dimensional supergravity theory
was first considered in \cite{sys}. Later many interesting solutions were
found including the novel black ring solutions with an event horizon of
topology $S^{1}\times S^{2}$\cite{ring}. Five-dimensional supergravity
theories with vector multiplets in all possible space-time signatures were
recently considered in \cite{euclidean, sig}. A rigorous analysis of
supersymmetry algebras in all space-time signatures was given in \cite{gt}.
A generalisation of the results of \cite{sys} related to time-like solutions
to all five-dimensional supergravity was considered in \cite{hyper}

In addition to the supergravity models based on symmetric scalar manifolds,
a large class of supergravities with Lorentzian signature can be obtained as
compactifications of eleven-dimensional supergravity \cite{cjs} on a
Calabi-Yau 3-folds \cite{cad}. Similarly, five-dimensional supergravity
theories in various space-time signatures can be obtained via Calabi-Yau
3-folds compactification of the exotic eleven-dimensional supergravity
theories constructed by Hull \cite{Hull}.

It is our purpose in the present work to study Kasner-like cosmological and
static solutions to five-dimensional supergravity theories. The analysis of
classical time-dependent gravitational solutions in string theory and
supergravities is important to investigate the role that these theories can
play in gravitational physics and cosmology.

The plan of this paper is as follows. In the next section we briefly review
the vacuum Kasner metrics in four and five dimensions and their extensions
to solutions of Einstein-Maxwell theory. In section three, we present some
basic properties of the ungauged five-dimensional $N=2$ supergravity theory
and the very special geometry underlying its structure as well as the
equations of motion for the metric, gauge and scalar fields of the theory.
We present two classes of solutions with some explicit examples in section
four where we also include a summary of our results.

\section{Solutions of Einstein-Maxwell is 4 and 5 dimensions}

The original vacuum four-dimensional Euclidean Kasner solution depends only
on one variable and is given by \cite{kasner} 
\begin{equation}
ds^{2}=x_{1}^{2a_{1}}dx_{1}^{2}+x_{1}^{2a_{2}}dx_{2}^{2}+x_{1}^{2a_{3}}dx_{3}^{2}+x_{1}^{2a_{4}}dx_{4}^{2}%
\text{ },
\end{equation}%
where the constants $a_{i}$, known as the Kasner exponents, satisfy two
conditions 
\begin{equation}
a_{2}+a_{3}+a_{4}=1+a_{1},\text{ \ \ \ \ }a_{2}^{2}+a_{3}^{2}+a_{4}^{2}=%
\left( 1+a_{1}\right) ^{2}.  \notag
\end{equation}%
The Kasner metric is actually valid for all space-time signatures \cite%
{harvey} and with a change of coordinates can take the form 
\begin{equation}
ds^{2}=\epsilon _{0}d\tau ^{2}+\left( \epsilon _{1}\tau ^{2a}dx^{2}+\epsilon
_{2}\tau ^{2b}dy^{2}+\epsilon _{3}\tau ^{2c}dz^{2}\right)  \label{gen}
\end{equation}%
where $\epsilon _{i}$ take the values $\pm 1$ and the constants satisfy the
conditions 
\begin{equation}
a+b+c=a^{2}+b^{2}+c^{2}=1.
\end{equation}%
Kasner metric is related to the metrics found by Weyl \cite{weyl},
Levi-Civita \cite{levi} and Wilson \cite{wilson} and was also rediscovered
by many authors \cite{harvey2}. For solutions with Lorentzian signature, we
can obtain the vacuum cosmological solutions%
\begin{equation}
ds^{2}=-d\tau ^{2}+\tau ^{2a}dx^{2}+\tau ^{2b}dy^{2}+\tau ^{2c}dz^{2},
\end{equation}%
as well as the anisotropic vacuum domain wall solutions 
\begin{equation}
ds^{2}=-r^{2a}dt^{2}+\left( dr^{2}+r^{2b}dy^{2}+r^{2c}dz^{2}\right) .
\end{equation}%
The vacuum Kasner metric can be promoted to solutions with a non-trivial
gauge field \cite{kbranes, kt}. For Einstein-Maxwell theory with the action 
\begin{equation}
S=\int d^{4}x\sqrt{\left\vert g\right\vert }\left( R-\frac{1}{4}%
\,F^{2}\right) ,
\end{equation}%
one obtains the solutions 
\begin{equation}
ds^{2}=\left( 1+\frac{\epsilon _{3}Q^{2}}{16c^{2}}\tau ^{2c}\right)
^{2}\left( \epsilon _{0}d\tau ^{2}+\epsilon _{1}\mathrm{\tau }%
^{2a}dx^{2}+\epsilon _{2}\mathrm{\tau }^{2b}dy^{2}\right) +\left( 1+\frac{%
\epsilon _{3}Q^{2}}{16c^{2}}\tau ^{2c}\right) ^{-2}\epsilon _{3}\mathrm{\tau 
}^{2c}dz^{2},  \notag
\end{equation}%
with the gauge field-strength two-form given by

\begin{equation}
F=Q\mathrm{\tau }^{2c-1}\left( 1+\frac{\epsilon _{3}Q^{2}}{16c^{2}}\tau
^{2c}\right) ^{-2}d\tau \wedge dz.
\end{equation}%
The dual magnetic Lorentzian cosmological solutions, with $\tau $ being a
time-like coordinate and $F=Qdx_{1}\wedge dx_{2}$ \cite{kt} are equivalent
to the family of anisotropic cosmologies of Rosen \cite{rosen}.

In five dimensional gravity with arbitrary space-time signature, the Kasner
metrics take the form 
\begin{equation}
ds^{2}=\epsilon _{0}d\tau ^{2}+\dsum\limits_{i=1}^{4}\epsilon _{i}\tau
^{2a_{i}}dx_{i}^{2}  \label{kas}
\end{equation}%
with the conditions 
\begin{equation}
\dsum\limits_{i=1}^{4}a_{i}=\dsum\limits_{i=1}^{4}a_{i}^{2}=1.  \label{f}
\end{equation}%
For Einstein-Maxwell action in five dimensions 
\begin{equation}
S=\int d^{5}x\sqrt{\left\vert g\right\vert }\left( R-\frac{\epsilon }{4}%
\,\,F^{2}\right)  \label{gact}
\end{equation}%
with $\epsilon =\pm 1,$ two classes of solutions can be found \cite{kbranes}%
. One is given by 
\begin{eqnarray}
ds^{2} &=&\left( c_{1}+\frac{\epsilon \epsilon _{4}Q^{2}\,}{12a_{4}^{2}c_{1}}%
\tau ^{2a_{4}}\right) \left( \epsilon _{0}d\tau
^{2}+\dsum\limits_{i=1}^{3}\epsilon _{i}\tau ^{2a_{i}}dx_{i}^{2}\right)
+\epsilon _{4}\left( c_{1}+\frac{\epsilon \epsilon _{4}Q^{2}\,}{%
12a_{4}^{2}c_{1}}\tau ^{2a_{4}}\right) ^{-2}\tau ^{2a_{4}}dx_{4}^{2}\text{ },
\notag \\
F &=&Q\,\tau ^{2a_{4}-1}\left( c_{1}+\frac{\epsilon \epsilon _{4}Q^{2}\,}{%
12a_{4}^{2}c_{1}}\tau ^{2a_{4}}\right) ^{-2}d\tau \wedge dx_{4},  \notag
\end{eqnarray}%
and the other class of solutions is for $F=Pdx_{1}\wedge dx_{2},$ with
constant $P$ and is given by%
\begin{equation}
ds^{2}=e^{2U}\left( \epsilon _{0}d\tau ^{2}+\epsilon _{1}\tau
^{2a_{1}}dx_{1}^{2}+\epsilon _{2}\tau ^{2a_{2}}dx_{2}^{2}\right)
+e^{-U}\left( \epsilon _{3}\tau ^{2a_{3}}dx_{3}^{2}+\epsilon _{4}\tau
^{2a_{4}}dx_{4}^{2}\right) \text{ },
\end{equation}%
with 
\begin{equation}
e^{U}=\left( c_{1}-\frac{\epsilon \epsilon _{0}\epsilon _{1}\epsilon
_{2}P^{2}}{12\left( a_{3}+a_{4}\right) ^{2}c_{1}}\tau ^{2\left(
a_{3}+a_{4}\right) }\right) .
\end{equation}

\section{$5D$ $N=2$ Supergravity and Very Special Geometry}

Ignoring the hypermultiplets, the Lagrangian of the bosonic fields of all
the five-dimensional supergravity can be given by \cite{GST, sig} 
\begin{equation}
\mathcal{L}_{5}=\sqrt{|g|}\left( R-\frac{\epsilon }{2}G_{IJ}F_{\mu \nu
}^{I}F^{\mu \nu J}-g_{ij}\partial _{\mu }\phi ^{i}\partial ^{\mu }\phi ^{j}-%
\frac{1}{24}\epsilon ^{\mu \nu \rho \sigma \tau }C_{IJK}F_{\mu \nu
}^{I}F_{\rho \sigma }^{J}A_{\tau }^{K}\right) .  \label{aa}
\end{equation}%
The indices $I,$ $J,$ $K$ take values $0,1,...,n,$ where $n$ is the number
of the scalar fields $\phi ^{i}.$ The constants $C_{IJK}$ are symmetric on
all indices. Note that for supergravity theories originating from Calabi-Yau
compactification, $C_{IJK}$ are the intersection numbers. The bosonic sector
of the theory can be described in terms of the so called very special
geometry \cite{speone}. We define the special coordinates $X^{I}=X^{I}(\phi )
$ satisfying 
\begin{equation}
X^{I}X_{I}=1,\qquad \mathcal{V=}{\frac{1}{6}}C_{IJK}X^{I}X^{J}X^{K}=1,
\label{cn}
\end{equation}%
where, $X_{I}$, the dual coordinate, is defined by 
\begin{equation}
X_{I}={\frac{1}{6}}C_{IJK}X^{J}X^{K}.  \label{d}
\end{equation}%
The gauge coupling metric is derived from the prepotential $\mathcal{V}$ and
is given by%
\begin{equation}
G_{IJ}=-{\frac{1}{2}}{\frac{\partial }{\partial X^{I}}}{\frac{\partial }{%
\partial X^{J}}}(\ln \mathcal{V})|_{\mathcal{V}=1}={\frac{9}{2}}X_{I}X_{J}-{%
\frac{1}{2}}C_{IJK}X^{K}.  \notag
\end{equation}%
Moreover, the scalar metric satisfies 
\begin{equation}
g_{ij}=G_{IJ}\partial _{i}X^{I}\partial _{j}X^{J}|_{\mathcal{V}=1}\text{ },
\label{gc}
\end{equation}%
where we have $\partial _{i}={\frac{\partial }{\partial \phi ^{i}}.}$ As
such, we have the relation $g_{ij}\partial _{\mu }\phi ^{i}\partial ^{\mu
}\phi ^{j}=G_{IJ}\partial _{\mu }X^{I}\partial ^{\mu }X^{J}.$ We also have
the very special geometry relations%
\begin{equation}
G_{IJ}X^{J}={\frac{3}{2}}X_{I}\,,\qquad dX_{I}\,=-\frac{2}{3}G_{IJ}dX^{J}.
\label{vsr}
\end{equation}%
The Einstein, Maxwell and scalar field equations of motion derived from (\ref%
{aa}) can be written as 
\begin{eqnarray}
R_{\mu \nu }-G_{IJ}\partial _{\mu }X^{I}\partial _{\nu }X^{J}-\epsilon {G}%
_{IJ}\left( F_{\mu \alpha }F_{\nu }{}^{\alpha }-{\frac{1}{6}}g_{\mu \nu
}F^{2}\right)  &=&0,  \label{sf} \\
\nabla _{\mu }\left( \epsilon G_{IJ}F^{I\mu \nu }\right) +\frac{1}{16}%
\epsilon ^{\nu \lambda \sigma \mu \tau }C_{IJK}F_{\lambda \sigma }^{J}F_{\mu
\tau }^{K} &=&0,  \label{gf} \\
\sqrt{|g|}\partial _{i}G_{IJ}\left( \frac{\epsilon }{2}F_{\mu \nu
}^{I}F^{\mu \nu J}+\partial _{\mu }X^{I}\partial ^{\mu }X^{J}\right)
-2\partial _{\mu }\left( \sqrt{|g|}G_{IJ}\partial ^{\mu }X^{J}\right)
\partial _{i}X^{I} &=&0.  \label{lo}
\end{eqnarray}

\section{General Solutions and Examples}

To solve for the equations of motion (\ref{sf})-(\ref{lo}), we start with
the following metric 
\begin{equation}
ds^{2}=e^{2U}\left( \epsilon _{0}d\tau ^{2}+\epsilon _{1}\mathrm{\tau }%
^{2a_{1}}dx^{2}+\epsilon _{2}\mathrm{\tau }^{2a_{2}}dy^{2}+\epsilon _{3}%
\mathrm{\tau }^{2a_{3}}dz^{2}\right) +\epsilon _{4}e^{-4U}\mathrm{\tau }%
^{2a_{4}}dw^{2}  \label{fc}
\end{equation}%
where the constants $a_{i},$ satisfying the Kasner conditions (\ref{f}) and $%
U$ is a function of $\tau $ only. For the gauge fields, we consider
solutions with only $F_{\tau w}^{I}$ non-vanishing and functions of $\tau ,$
the equations of motion (\ref{gf}) then reduce to 
\begin{equation*}
\partial _{\tau }\left( \tau e^{2U}G_{IJ}F^{J\tau w}\right) =0
\end{equation*}%
and thus imply the solution%
\begin{equation}
G_{IJ}F^{J\tau w}=e^{-2U}\frac{q_{I}}{\tau }.  \label{ge}
\end{equation}%
where $q_{I}$ being constants. Substituting the value of the gauge
field-strength into the Einstein equations of motion (\ref{sf}), we obtain

\begin{eqnarray}
\ddot{U}+2\dot{U}^{2}+\left( 1-2a_{4}\right) \frac{\dot{U}}{\tau } &=&-\frac{%
1}{3}G_{IJ}\dot{X}^{I}\dot{X}^{J},  \label{eone} \\
3\left( \ddot{U}+\frac{\dot{U}}{\tau }\right) &=&\epsilon \epsilon _{4}\tau
^{2a_{4}-2}e^{-4U}G^{IJ}q_{I}q_{J}\text{ },  \label{etwo}
\end{eqnarray}%
here the dot symbol indicates differentiation with respect to the coordinate 
$\tau .$ In order to solve these equations, we write the dual coordinate in
the form 
\begin{equation}
X_{I}(\tau )=\frac{1}{3}e^{-2U}f_{I}(\tau ),  \label{sa}
\end{equation}%
then with the help of the relations of very special geometry, the following
equations can be derived 
\begin{eqnarray}
\dot{X}_{I} &=&-2\dot{U}X_{I}+\frac{1}{3}e^{-2U}\dot{f}_{I}\text{ },  \notag
\\
\dot{U} &=&\frac{1}{6}e^{-2U}\dot{f}_{I}X^{I},  \notag \\
\ddot{U} &=&\frac{1}{6}e^{-2U}\left( X^{I}\ddot{f}_{I}-\frac{1}{2}%
e^{-2U}G^{IJ}\dot{f}_{I}\dot{f}_{J}\right) ,  \notag \\
G_{IJ}\dot{X}^{I}\dot{X}^{J} &=&-3\ddot{U}-6\dot{U}^{2}+\frac{1}{2}%
e^{-2U}X^{I}\ddot{f}_{I}\text{ },
\end{eqnarray}%
which when substituted in the equations (\ref{eone}) and (\ref{etwo}) give%
\begin{eqnarray}
X^{I}\left[ \ddot{f}_{I}+\left( 1-2a_{4}\right) \frac{\dot{f}_{I}}{\tau }%
\right] &=&0,  \label{fee} \\
e^{-2U}X^{I}\left( \ddot{f}_{I}+\frac{\dot{f}_{I}}{\tau }\right)
-e^{-4U}G^{IJ}\left( \frac{1}{2}\dot{f}_{I}\dot{f}_{J}+2\epsilon \epsilon
_{4}q_{I}q_{J}\tau ^{2a_{4}-2}\right) &=&0.  \label{see}
\end{eqnarray}%
Equation (\ref{fee}) admits the solution%
\begin{equation}
f_{I}=\left( A_{I}+B_{I}\mathrm{\tau }^{2a_{4}}\right)
\end{equation}%
with constant $A_{I}$ and $B_{I}.$ Equation (\ref{see}) then implies the
conditions

\begin{equation}
G^{IJ}\left( A_{J}B_{I}a_{4}^{2}-\epsilon \epsilon _{4}q_{I}q_{J}\right) =0.
\label{conone}
\end{equation}%
The scalar equation of motion (\ref{lo}) for our solutions reduces to 
\begin{equation}
\partial _{i}G^{IJ}\left( {\frac{1}{2}}\dot{f}_{I}\dot{f}_{J}-\frac{1}{2}%
f_{J}\left( \frac{\dot{f}_{I}}{\tau }+\ddot{f}_{I}\right) +{2\epsilon
_{4}\epsilon }q_{I}q_{J}\mathrm{\tau }^{2a_{4}-2}\right) =0
\end{equation}%
which implies the conditions 
\begin{equation}
\partial _{i}G^{IJ}\left( A_{J}B_{I}a_{4}^{2}-\epsilon \epsilon
_{4}q_{I}q_{J}\right) =0.  \label{contwo}
\end{equation}%
Note that (\ref{cn}) implies the relation 
\begin{equation}
e^{4U}=\frac{1}{6}G^{IJ}f_{I}f_{J}.
\end{equation}%
We consider a second class of of solutions represented by the metric 
\begin{equation}
ds^{2}=e^{2U}\left( \epsilon _{0}d\tau ^{2}+\epsilon _{1}\mathrm{\tau }%
^{2a_{1}}dx^{2}+\epsilon _{2}\mathrm{\tau }^{2a_{2}}dy^{2}\right)
+e^{-U}\left( \epsilon _{3}\mathrm{\tau }^{2a_{3}}dz^{2}+\epsilon _{4}%
\mathrm{\tau }^{2a_{4}}dw^{2}\right)  \label{sc}
\end{equation}%
and the gauge field-strength two-form $F^{I}=p^{I}dx\wedge dy$ with constant 
$p^{I}.$ For the metric (\ref{sc}), one obtains for the Einstein equations
of motion 
\begin{eqnarray}
\ddot{U}+\left( 1-2l\right) \frac{\dot{U}}{\tau }+\dot{U}^{2} &=&-\frac{2}{3}%
G_{IJ}\dot{X}^{I}\dot{X}^{J},  \label{ri1} \\
\ddot{U}+\frac{\dot{U}}{\tau } &=&-\frac{2}{3}\epsilon \epsilon _{0}\epsilon
_{1}\epsilon _{2}e^{-2U}\tau ^{-2s}{G}_{IJ}p^{I}p^{J}  \label{ri2}
\end{eqnarray}%
where we have $l=a_{3}+a_{4}$ and $s=a_{1}+a_{2}.$ To proceed in finding
solutions, we set 
\begin{equation}
X^{I}(\tau )=e^{-U}h^{I}(\tau ),
\end{equation}%
then using the relations of very special geometry, we obtain the following
relations%
\begin{eqnarray}
\dot{U} &=&e^{-U}X_{I}\dot{h}^{I},  \notag \\
\ddot{U} &=&-\frac{2}{3}e^{-2U}G_{IJ}\dot{h}^{I}\dot{h}^{J}+e^{-U}X_{I}\ddot{%
h}^{I},  \notag \\
G_{IJ}\dot{X}^{I}\dot{X}^{J} &=&-\frac{3}{2}\dot{U}^{2}+e^{-2U}G_{IJ}\dot{h}%
^{I}\dot{h}^{J}.  \label{cs}
\end{eqnarray}%
Using (\ref{cs}), equations (\ref{ri1}) and (\ref{ri2}) then give 
\begin{eqnarray}
X_{I}\left[ \ddot{h}^{I}+\left( 1-2l\right) \frac{\dot{h}^{I}}{\tau }\right]
&=&0, \\
G_{IJ}\left[ \dot{h}^{I}\dot{h}^{J}-h^{J}\left( \ddot{h}^{I}+\frac{\dot{h}%
^{I}}{\tau }\right) -\epsilon \epsilon _{0}\epsilon _{1}\epsilon _{2}\tau
^{-2s}p^{I}p^{J}\right] &=&0,
\end{eqnarray}%
\ which can be solved by%
\begin{equation}
h^{I}=C^{I}+D^{I}\tau ^{2l},
\end{equation}%
with the conditions 
\begin{equation}
G_{IJ}\left( 4l^{2}C^{J}D^{I}+\epsilon \epsilon _{0}\epsilon _{1}\epsilon
_{2}p^{I}p^{J}\right) =0.
\end{equation}%
The scalar equation of motion also implies the conditions 
\begin{equation}
\partial _{i}G_{IJ}\left( 4l^{2}C^{J}D^{I}+\epsilon \epsilon _{0}\epsilon
_{1}\epsilon _{2}p^{I}p^{J}\right) =0.
\end{equation}%
Moreover, the condition (\ref{cn}) for our solutions implies%
\begin{equation}
e^{3U}={\frac{1}{6}}C_{IJK}h^{I}h^{J}h^{K}.
\end{equation}

In what follows, some explicit solutions are constructed. We start with the
supergravity models where the scalar manifold is a symmetric space. For such
theories, the prepotential $\mathcal{V}$ \ defined in (\ref{cn}) is related
to the norm forms of degree three Euclidean Jordan algebras. For non-simple
Jordan algebras, the corresponding symmetric scalar manifolds $\mathcal{M}$
are given by%
\begin{equation}
\mathcal{M}=SO(1,1)\times \frac{SO(n-1,1)}{SO(n-1)}  \label{scalman}
\end{equation}%
and the prepotential ${\mathcal{V}}$ factorizes into a linear times a
quadratic form in $(n-1)$ scalars and takes the form 
\begin{equation}
{\mathcal{V}}={\frac{1}{2}}X^{1}\left( \eta _{ab}X^{a}X^{b}\right) ,\quad
a,b=2,\dots ,n.
\end{equation}%
For these models, the following relations hold 
\begin{equation}
X^{I}={\frac{9}{2}}C^{IJK}X_{J}X_{K}\text{ },\text{ \ \ \ }%
G^{IJ}=2X^{I}X^{J}-6C^{IJK}X_{K}\text{ },
\end{equation}%
where $C^{IJK}$ is defined by 
\begin{equation}
C^{IJK}=\delta ^{II^{\prime }}\delta ^{JJ^{\prime }}\delta ^{KK^{\prime
}}C_{I^{\prime }J^{\prime }K^{\prime }}\,.
\end{equation}%
Explicitly, the components of $G_{IJ}$ and its inverse are given by

\begin{eqnarray}
G_{11} &=&{\frac{9}{2}}\left( X_{1}\right) ^{2},\text{ \ \ \ }G_{1a}=0,\text{
\ \ \ }G_{ab}=\frac{9}{2}X_{a}X_{b}-\frac{1}{2}\eta _{ab}X^{1},  \notag \\
G^{11} &=&2\left( X^{1}\right) ^{2},\text{ \ \ \ \ \ }G^{1a}=0,\text{ \ \ \ }%
G^{ab}=2X^{a}X^{b}-6\eta ^{ab}X_{1}.
\end{eqnarray}%
For the first class of solutions in (\ref{fc}), the non-vanishing gauge
fields and scalars are given by 
\begin{eqnarray}
F^{1\tau w} &=&e^{-2U}G^{11}\frac{q_{1}}{\tau },\text{ \ \ \ \ \ \ \ \ \ \ \
\ \ \ \ \ }  \notag \\
F^{a\tau w} &=&e^{-2U}G^{ab}\frac{q_{b}}{\tau },  \notag \\
X_{1} &=&\frac{1}{3}e^{-2U}\left( A_{1}+B_{1}\mathrm{\tau }^{2a_{4}}\right) ,%
\text{ \ \ \ }  \notag \\
X_{a} &=&\frac{1}{3}e^{-2U}\left( A_{a}+B_{a}\mathrm{\tau }^{2a_{4}}\right) ,
\notag \\
\bigskip X^{1} &=&{\frac{1}{2}}\eta ^{ab}e^{-4U}\left( A_{a}+B_{a}\mathrm{%
\tau }^{2a_{4}}\right) \left( A_{b}+B_{b}\mathrm{\tau }^{2a_{4}}\right) ,%
\text{ \ }  \notag \\
X^{a} &=&\eta ^{ab}e^{-4U}\left( A_{1}+B_{1}\mathrm{\tau }^{2a_{4}}\right)
\left( A_{b}+B_{b}\mathrm{\tau }^{2a_{4}}\right) .
\end{eqnarray}
Note that for these models we have%
\begin{equation*}
X^{1}X_{1}=\frac{1}{3},\text{ \ \ \ }X^{a}X_{a}=\frac{2}{3},
\end{equation*}%
thus we obtain for our solution 
\begin{equation}
e^{6U}={\frac{1}{2}}\eta ^{ab}\left( A_{a}+B_{a}\mathrm{\tau }%
^{2a_{4}}\right) \left( A_{b}+B_{b}\mathrm{\tau }^{2a_{4}}\right) \left(
A_{1}+B_{1}\mathrm{\tau }^{2a_{4}}\right) .
\end{equation}

The conditions (\ref{conone}) and (\ref{contwo}) are satisfied by allowing
for two independent charges $q_{1}$ and $q_{a}=Q,$ and $A_{a}=B_{a}=A,$ with 
\begin{equation}
\left( q_{1}\right) ^{2}=\epsilon \epsilon _{4}A_{1}B_{1}a_{4}^{2},\text{ \
\ }Q^{2}=\epsilon \epsilon _{4}A^{2}a_{4}^{2}.
\end{equation}%
Similarly one can construct solutions with two independent charges for the
second class of metrics given in (\ref{sc}).

Next we consider the $U(1)^{3}$ supergravity with the prepotential $\mathcal{%
V}=X^{1}X^{2}X^{3}=1$ and $G_{IJ}=\frac{1}{2\left( X^{I}\right) ^{2}}\delta
_{IJ}.$ We first consider the first class of solutions given in (\ref{fc}).
Using (\ref{sa}), we obtain

\begin{eqnarray}
X_{1} &=&{\frac{1}{3}}X^{2}X^{3}=\frac{1}{3}e^{-2U}H_{1},\text{ \ \ \ \ \ }
\\
X_{2} &=&{\frac{1}{3}}X^{1}X^{3}=\frac{1}{3}e^{-2U}H_{2},\text{ \ \ \ \ \ \ }
\\
X_{3} &=&{\frac{1}{3}}X^{1}X^{2}=\frac{1}{3}e^{-2U}H_{3},
\end{eqnarray}%
where $H_{I}=\left( a_{I}+b_{I}\mathrm{\tau }^{2a_{4}}\right) .$ The special
coordinates and the metric are given by 
\begin{equation}
X^{I}=\frac{\left( H_{1}H_{2}H_{3}\right) ^{1/3}}{H_{I}},\text{ \ \ }
\label{ss}
\end{equation}%
and 
\begin{equation}
ds^{2}=\left( H_{1}H_{2}H_{3}\right) ^{1/3}\left( \epsilon _{0}d\tau
^{2}+\epsilon _{1}\mathrm{\tau }^{2a_{1}}dx^{2}+\epsilon _{2}\mathrm{\tau }%
^{2a_{2}}dy^{2}+\epsilon _{3}\mathrm{\tau }^{2a_{3}}dz^{2}\right) +\epsilon
_{4}\left( H_{1}H_{2}H_{3}\right) ^{-2/3}\mathrm{\tau }^{2a_{4}}dw^{2}.
\label{stue}
\end{equation}%
Using (\ref{ge}), we obtain for the gauge fields (with no summation over the
index $I$)%
\begin{equation}
F^{I\tau w}=\frac{2q_{I}}{\tau H_{I}^{2}}\left( H_{1}H_{2}H_{3}\right) ^{1/3}
\label{gs}
\end{equation}%
with the three independent charges given by $q_{I}^{2}=\epsilon \epsilon
_{4}a_{I}b_{I}a_{4}^{2}.$

Using the general class of solutions given in (\ref{sc}), we obtain for the $%
U(1)^{3}$ supergravity the solutions given by 
\begin{equation}
ds^{2}=\left( H_{1}H_{2}H_{3}\right) ^{2/3}\left( \epsilon _{0}d\tau
^{2}+\epsilon _{1}\mathrm{\tau }^{2a_{1}}dx^{2}+\epsilon _{2}\mathrm{\tau }%
^{2a_{2}}dy^{2}\right) +\left( H_{1}H_{2}H_{3}\right) ^{-1/3}\left( \epsilon
_{3}\mathrm{\tau }^{2a_{3}}dz^{2}+\epsilon _{4}\mathrm{\tau }%
^{2a_{4}}dw^{2}\right)   \label{stum}
\end{equation}%
and 
\begin{equation}
F^{I}=q^{I}dx\wedge dy,\text{ \ \ \ \ \ }X^{I}=\frac{H^{I}}{\left(
H_{1}H_{2}H_{3}\right) ^{1/3}},\text{ }  \label{sss}
\end{equation}%
where 
\begin{equation*}
H^{I}=A^{I}+B^{I}\tau ^{2l}
\end{equation*}%
and the three independent charges are given by 
\begin{equation}
\left( q^{I}\right) ^{2}=-4\epsilon \epsilon _{0}\epsilon _{1}\epsilon
_{2}l^{2}A^{I}B^{I}.
\end{equation}

As as special case, we can obtain from (\ref{stue}) a Lorentzian solution
for which all Kasner exponents vanish except for $a_{4}.$ This takes the form

\begin{equation}
ds^{2}=\left( H_{1}H_{2}H_{3}\right) ^{1/3}\left( -d\tau
^{2}+dx^{2}+dy^{2}+dz^{2}\right) +\left( H_{1}H_{2}H_{3}\right) ^{-2/3}\tau
^{2}dw^{2},
\end{equation}%
where the gauge fields and scalars are as in (\ref{ss}) and (\ref{gs}) and
we choose

\begin{equation}
H_{I}=\left( 1+\epsilon q_{I}^{2}\tau ^{2}\right) .
\end{equation}%
This solution can be thought of as a Melvin cosmology with non-trivial
scalar fields and three independent electric charges. Another special
solution can be obtained by analytic continuation or by simply setting $%
\epsilon _{0}=-\epsilon _{1}=\epsilon _{2}=\epsilon _{3}=\epsilon
_{4}=a_{4}=1$ in (\ref{stue}). This can be given (after relabelling of the
coordinates) by 
\begin{equation}
ds^{2}=\left( H_{1}H_{2}H_{3}\right) ^{1/3}\left(
-dt^{2}+dr^{2}+dy^{2}+dz^{2}\right) +\left( H_{1}H_{2}H_{3}\right)
^{-2/3}r^{2}dw^{2}
\end{equation}%
with 
\begin{eqnarray}
H_{I} &=&\left( 1+\epsilon q_{I}^{2}r^{2}\right) ,  \notag \\
X^{I} &=&\frac{\left( H_{1}H_{2}H_{3}\right) ^{1/3}}{H_{I}},\text{ \ \ \ \ }%
F^{Irw}=\frac{2q_{I}}{rH_{I}^{2}}\left( H_{1}H_{2}H_{3}\right) ^{1/3}.
\end{eqnarray}%
This solution can be thought of an anisotropic Melvin domain wall with
non-trivial gauge and scalar fields. Solutions with cyclic $w$ correspond to
five dimensional generalisations of Melvin fluxtubes \cite{melvin}. One can
also obtain from (\ref{stum}), magnetically charged cosmological solutions
represented by 
\begin{eqnarray}
ds^{2} &=&\left( H_{1}H_{2}H_{3}\right) ^{2/3}\left( -d\tau
^{2}+dx^{2}+dy^{2}\right) +\left( H_{1}H_{2}H_{3}\right) ^{-1/3}\left(
dz^{2}+\mathrm{\tau }^{2}dw^{2}\right)   \notag \\
H^{I} &=&1+\epsilon \frac{\left( q^{I}\right) ^{2}}{4}\tau ^{2},  \label{ssr}
\end{eqnarray}%
with bosonic fields as given in (\ref{sss}). Static solutions can also be
obtained from (\ref{ssr}) by analytic continuation.

It is known that any solution of the five-dimensional $U(1)^{3}$
supergravity can be uplifted to a solution of eleven-dimensional
supergravity and a solution of type IIB supergravity. \ The
eleven-dimensional solutions take the form \cite{ring}

\begin{equation}
ds_{11}^{2}=ds_{5}^{2}+X^{1}\left( dz_{1}^{2}+dz_{2}^{2}\right) +X^{2}\left(
dz_{3}^{2}+dz_{4}^{2}\right) +X^{3}\left( dz_{5}^{2}+dz_{6}^{2}\right)
\end{equation}%
where $ds_{5}^{2}$ represents the five-dimensional metric and the
coordinates $z^{i}$, $i=1,..,6$, parametrize $T^{6}.$ The type IIB solution
is given by 
\begin{equation}
ds_{10}^{2}=\left( X^{3}\right) ^{1/2}ds_{5}^{2}+\left( X^{3}\right)
^{-3/2}\left( dz+A^{3}\right) ^{2}+X^{1}\left( X^{3}\right) ^{1/2}\left(
dz_{1}^{2}+dz_{2}^{2}+dz_{3}^{2}+dz_{4}^{2}\right)
\end{equation}%
with the dilaton $\phi $ and the Ramond-Ramond 3-form field strength $%
F_{(3)} $ given by

\begin{equation}
e^{2\phi }=\frac{X^{1}}{X^{2}},\text{ \ \ \ }F_{(3)}=\left( X^{1}\right)
^{-2}\ast _{5}F^{1}+F^{2}\wedge \left( dz+A^{3}\right) ,
\end{equation}%
with $\ast _{5}$ being the Hodge dual with respect to $ds_{5}^{2}.$

In conclusion, time-dependent and static gravitational solutions for the
theory of five-dimensional supergravity with non-trivial scalar and gauge
fields in various space-time dimensions have been considered. As in the
original metric constructed by Kasner, our solutions depend only on one
parameter which we have denoted by $\tau .$ Two classes of solutions were
found and some explicit solutions with two and three charges were
constructed. Depending on the choice of coordinates, the resulting solutions
can be thought of as generalizations of the electric Melvin or magnetic
Rosen cosmologies, flux-branes and domain walls. Our results can be
generalised to supergravity with various space-time dimensions. Another
interesting direction for further investigation is to consider solutions of
gauged supergravity theories which have a cosmological constant or a scalar
potential. Those solutions will have potential applications to (A)dS/CFT
correspondence. We hope to report on this in our future work.

\bigskip

{\flushleft{\textbf{Acknowledgements:}}} This work is supported in part by
the National Science Foundation under grant number PHY-1620505.

\end{document}